\documentclass[reprint,amsmath,amssymb,showkeys,aps,prb]{revtex4-1}
\usepackage{natbib}
\usepackage{graphicx}
\usepackage{dcolumn}
\usepackage{bm}
\usepackage{hyperref}

\begin{document}
\preprint{APS/123-QED}
\title{Stick-Slip Nanofriction in Trapped Cold Ion Chains}
\author{D. Mandelli $^{1}$, A. Vanossi$^{2,1}$, and E. Tosatti$^{1,2,3}$}
\affiliation{
$^1$ International School for Advanced Studies (SISSA), 
     Via Bonomea 265, 34136 Trieste, Italy \\
$^2$ CNR-IOM Democritos National Simulation Center, 
     Via Bonomea 265, 34136 Trieste, Italy \\
$^3$ International Centre for Theoretical Physics (ICTP), 
     Strada Costiera 11, 34014 Trieste, Italy
            }
\date{\today}
\begin{abstract}
\section*{Abstract}
 Stick-slip -- the sequence of mechanical instabilities through which 
 a slider advances on a solid substrate -- is pervasive throughout sliding 
 friction, from nano to geological scales. Here we suggest that trapped 
 cold ions in an optical lattice can also be of help in understanding
 stick-slip friction, and also the way friction changes when one of the 
 sliders undergoes structural transitions. For that scope, we simulated 
 the dynamical properties of a 101-ions chain, driven to slide back and 
 forth by a slowly oscillating electric field in an incommensurate periodic 
 ``corrugation'' potential of increasing magnitude $U_0$. We found the 
 chain sliding to switch, as $U_0$ increases and before the Aubry transition,
 from a smooth-sliding regime with low dissipation to a stick-slip regime 
 with high dissipation. In the stick-slip regime the onset of overall 
 sliding is preceded by precursor events consisting of partial slips of 
 few ions only, leading to partial depinning of the chain, a nutshell 
 remnant of precursor events at the onset of motion also observed in 
 macroscopic sliders. Seeking to identify the possible effects on friction
 of a structural transition, we reduced the trapping potential aspect ratio 
 until the ion chain shape turned from linear to zigzag. Dynamic friction 
 was found to rise at the transition, reflecting the opening of newer 
 dissipation channels.
\end{abstract}
\keywords{}
\maketitle

\section{Introduction}
 Similarly to colloidal monolayers driven across laser-generated 
 surfaces~\cite{BOHL,MANI}, linear chains of cold ions trapped inside 
 optical lattices have been recently proposed as novel candidates for studies 
 in the field of friction~\cite{BEN}. One of the motivations has been the 
 possibility to observe, thanks to their exceptional parameter tunability, 
 the long theorized Aubry transition, namely the switch between a regular 
 frictional state and the ``superlubric''  state of vanishing static friction 
 between idealized incommensurate one dimensional (1D) ``crystals''. The key 
 feature of friction between solid bodies is hysteresis, that is the 
 difference between to and fro motion. In  time-periodic sliding motion 
 for example, hysteresis is responsible for the finite area enclosed by 
 the force-displacement cycle, which exactly equals the frictional
 heat per cycle. Smallest when the sliding regime is smooth, friction turns 
 large when sliding occurs by stick-slip -- a discontinuous stop and go which
 constitutes the largest and commonest  source of frictional hysteresis. 
 Generally triggered by mechanical instabilities, stick-slip takes place 
 at geological, ordinary, and at nanometer length scales 
 alike~\cite{BOWD,BOBOOK,VANRMP}. Restricting here to the nano and 
 microscale, which is the focus of much current work, we are naturally 
 interested in microscopical systems exhibiting a controlled transition 
 between smooth and stick-slip sliding regimes.
 
\begin{figure}[!t]
 \begin{center}
 \includegraphics[angle=0, width=0.5\textwidth]{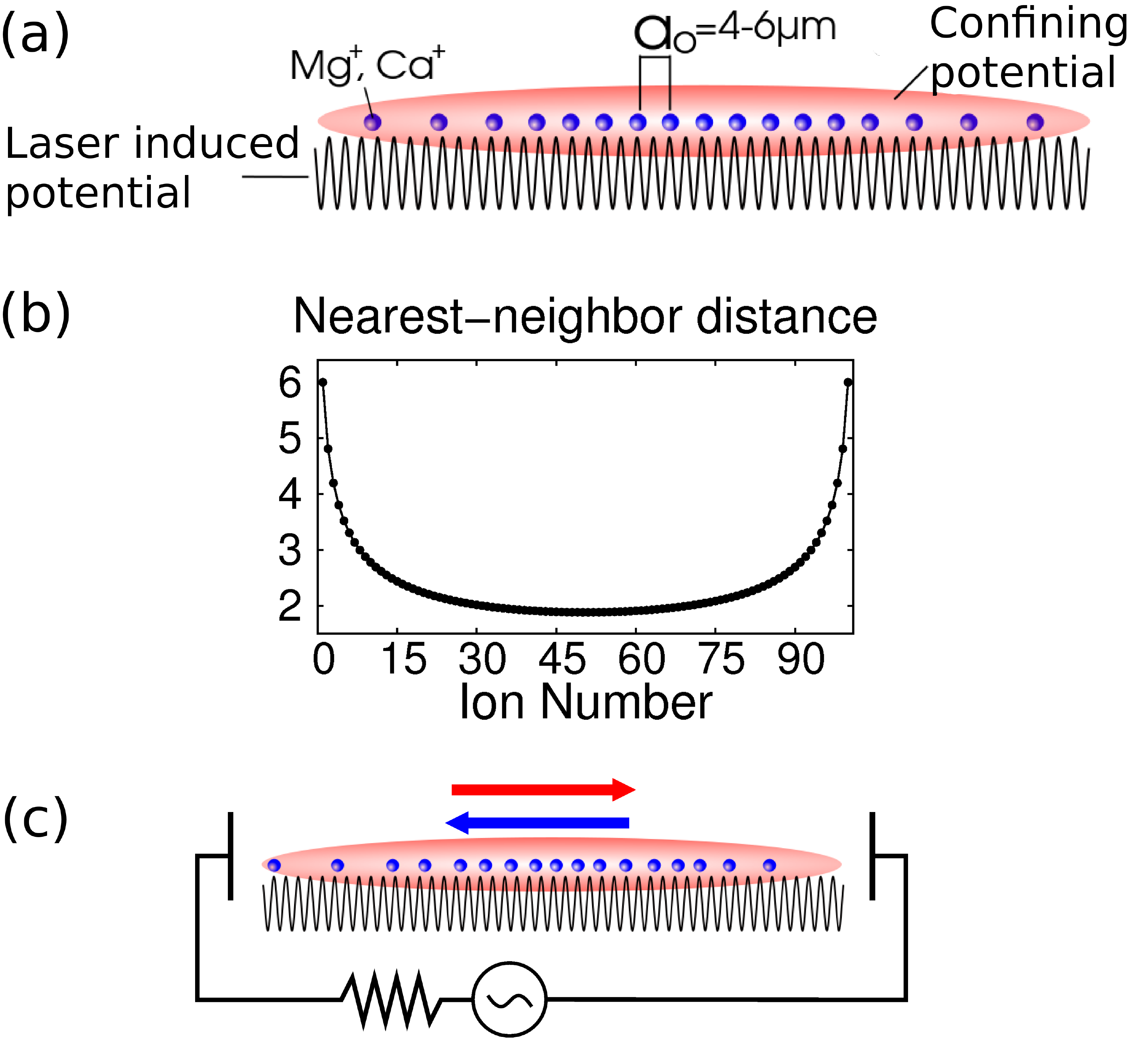}
 \end{center}
 \caption{(a) Schematic of a linear ion chain trapped by an anisotropic 
              confining potential. Typical ions used in experiments are 
              Ca$^+$ or Mg$^+$, the ion-ion distance at the center of the 
              chain is of the order of few $\mu m$.
          (b) Nearest-neighbor distance (in dimensionless units,
              see Section~\ref{sec1}) between the ions of a 101-ions
              chain at rest and in absence of the corrugation potential.
          (c) A possible experimental set up for the study of the dynamics 
              of the chain using an external oscillating electric field.
         }
 \label{fig1}
 \end{figure}

\begin{figure}[!t]
 \begin{center}
 \includegraphics[angle=0, width=0.5\textwidth]{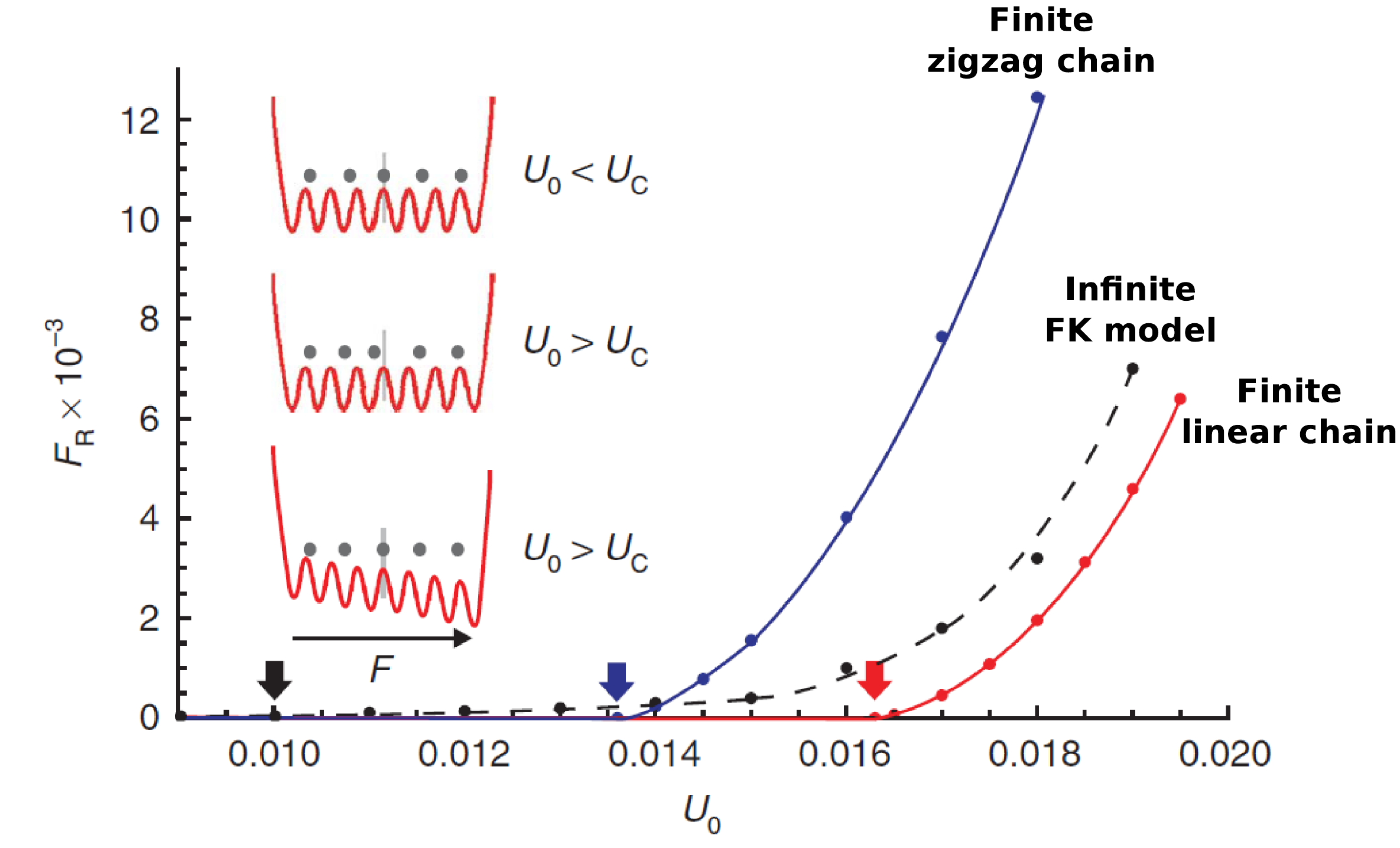}
 \end{center}
 \caption{ (Color online) Effective static friction of a 101-ions chain 
           in the linear (red) and zigzag (blue) configuration
           plotted against the corrugation potential amplitude $U_0$.
           The black dashed line is the static friction of the ideal
           FK model. (From Ref.~[\onlinecite{BEN}]) 
           }
 \label{fig2}
 \end{figure}

 One dimensional periodic (``crystalline'') sliding models, although 
 highly simplified, have long been used to illustrate frictional phenomena 
 between periodic lattices~\cite{VABR}. In the so-called Frenkel Kontorova 
 (FK) model~\cite{FK}, a harmonic chain of classical masses with average 
 spacing $a_o$ in a sinusoidal periodic potential of amplitude $U_0$ and 
 wavelength $\lambda$ (leading to a commensuration ratio $\eta$=$a_o/\lambda$
 between the two) idealizes the sliding of two crystalline surfaces. 
 Irrational values of $\eta$ characterize the most interesting incommensurate
 case between slider and substrate. Aubry~\cite{AUB1,AUB2} proved long ago that a
 transition (where the ground state ``hull function'' exhibits analyticity 
 breaking) occurs for increasing $U_0$, from what is now known as a 
 {\it superlubric} state where the static friction $F_S$ -- the minimal 
 force capable of initiating sliding -- is exactly zero, to a {\it pinned} 
 state where $F_S$ is finite. While exceptionally low friction between 
 incommensurate 3D surfaces has indeed been observed 
 experimentally~\cite{FRE}, there had not been so far experimental 
 demonstrations of the Aubry transition in genuinely 1D systems.
 Cold ion traps were recently invoked as possible candidates to display the 
 Aubry transition, thereby surprisingly entering the field of 
 nanotribology~\cite{BEN}. 
 Although not identical to the FK model, the
 physics of repulsive 1D particles is expected to be essentially the same
 as each particle can still be seen as occupying the center of some overall
 harmonic potential.
 Experimentally~\cite{HONG}, chains of up to 
 several tens of positive ions such as Ca$^+$ can be stabilized using rf 
 quadrupolar fields and cooled down to temperatures below 1 $\mathrm{\mu}$K.
 By tuning the confining cigar-shaped potential to a sufficiently elongated 
 form, the ions can be forced to form linear chains. The periodic optical 
 lattice potential for the ions is provided by a laser standing wave 
 (see Fig.~\ref{fig1}a).\\
 The confined ion chains do constitute 1D crystal segments, but are not 
 really homogeneous. The nearest neighbor ion-ion distance, fairly constant 
 at the center, increases at the periphery and diverges near the extremities,
 as shown in Fig.~\ref{fig1}b.
 Still, the chain center is a reasonable realization of an FK-like model, and
 some of the properties of an ideal infinite system can be in principle 
 realized and observed there. When $\lambda$ is incommensurate with respect 
 to the central ion-ion spacing $a_o$ one can achieve, according to our 
 recent predictions~\cite{BEN}, a strong and observable remnant of the Aubry transition 
 also in such trapped ion chains. In the confined ion chain, the standard 
 Aubry transition, which in the infinite chain occurs when the periodic 
 potential (``corrugation'') amplitude $U_0$ exceeds some critical 
 threshold $U_c$, is replaced by a static, symmetry breaking transition 
 of the ground state chain configuration and geometry. Benassi and 
 coworkers~\cite{BEN} proposed to observe this transition by measuring the 
 external uniform force $F_R$ needed to restore the symmetry. Simulations 
 indeed showed that the effective static friction force $F_R$ behaves and 
 grows as a function of $U_0 > U_c$ very closely like the static friction 
 force $F_S$ of the ideal infinite chain thus demonstrating the connection 
 between the two (see Fig.~\ref{fig2}).

 In this work we move from static friction to the dynamical sliding properties
 of the ion chain, once depinned by an additional external electric field, as
 a function of the periodic corrugation amplitude $U_0$
 (see Fig.~\ref{fig1}c). We show that the trapped cold ions can slide either
 smoothly or by stick-slip, with a parameter-controlled transition and a
 correspondingly strong frictional rise between the former and the latter.
 Drawing an analogy with macroscopic frictional experiments, the corrugation 
 $U_0$ plays here the role of the load in ordinary sliding friction. As
 expected, we find and characterize the transition from a poorly dissipative 
 smooth sliding regime to a highly dissipative stick-slip regime as $U_0$ is 
 increased. Moreover, since the cold ion ground state geometric configuration 
 can be pushed across parameter-driven structural transitions by changing the
 trapping potential conditions, we investigate the effect of a phase transition
 on sliding friction, which is of interest as well~\cite{BEN1}. As is known 
 both theoretically~\cite{SCHI,MOR3} and experimentally~\cite{RAIZ}, a change
 of aspect ratio in the confining trap effective potential causes the ion 
 chain to cross a series of structural transitions. If and when for a 
 sufficiently long chain these transitions can be considered continuous, the
 friction behavior near the transition point could show remnants of the 
 chain's critical behavior, as recently suggested theoretically~\cite{BEN1}.

 In anticipation of future experiments, we carried out classical molecular 
 dynamics simulations of a 101-ions chain sliding in a golden ratio 
 incommensurate corrugated potential, with a view to predict and discuss 
 the basic dynamic frictional phenomena of an electric field solicited
 trapped ion chain. In Section~\ref{sec1} we will describe the model 
 and the protocol used for the MD simulations. Section~\ref{sec2} will be 
 devoted to the resulting smooth to stick-slip frictional switch and the
 observation of precursor events at the onset of the chain sliding. In 
 Section~\ref{sec3} we describe the change of the frictional behavior 
 across the linear-zigzag structural transition. Finally Section~\ref{sec4}
 contains our discussion and conclusions.
\section{Trapped Ion Chain Model and Sliding Simulation Protocol}
\label{sec1}
\begin{figure}[!t]
 \begin{center}
 \includegraphics[angle=0, width=0.5\textwidth]{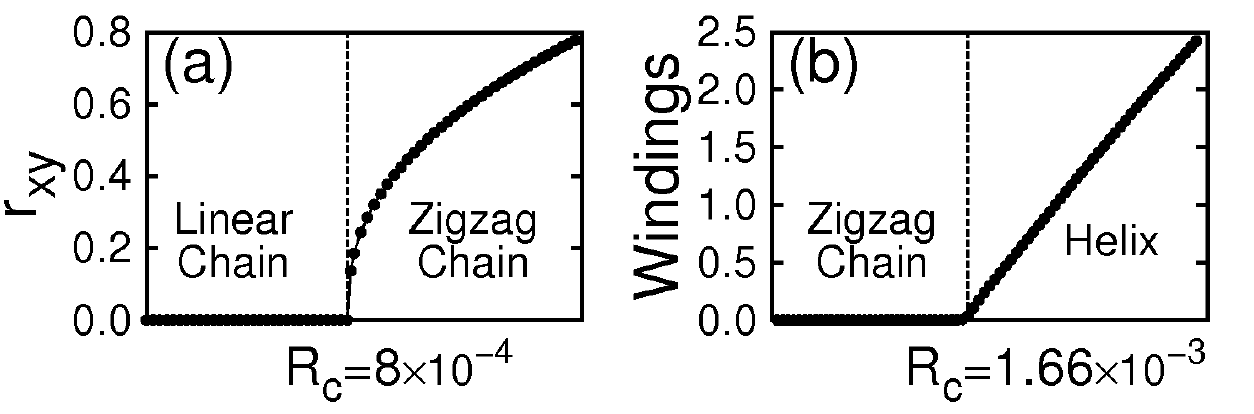}
 \end{center}
 \caption{(a) Linear-zigzag structural transition of a 101-ions chain. 
          $r_{xy}$ is the maximum displacement of the ions from the $z$ axis.
          (b) Zigzag-helix transition. The order parameter
          is the number of windings of the ions around the $z$ axis.}
 \label{fig3}
 \end{figure}

  The effective potential of an ion of charge $q$ in a linear anisotropic 
  (Paul) trap can be written as~\cite{TRAP}:
 \begin{equation}
 V_{eff}(x,y,z)=\frac{1}{2m}\left[ \omega_{\perp}^2(x^2+y^2)+
                \omega_{\parallel}^2z^2\right],
 \end{equation}
 where $m$ is the mass of the ion and $\omega_{\perp}^2$ and 
 $\omega_{\parallel}^2$ are the strengths of the confining effective potential,
 supposed to be harmonic, in the transverse and longitudinal directions. 
 In order to work in dimensionless units we define the length unit $d$:
 \begin{equation}
  d=\left(\frac{q^2}{4\pi \epsilon_0m\omega_{\perp}^2}\right)^{1/3}.
 \end{equation}
 We then measure masses in units of $m$, charges in units of $q$, energy in
 units of $q^2/(4\pi\epsilon_0d)$, forces in units of 
 $q^2/(4\pi\epsilon_0d^2)$ and time in units of $1/\omega_{\perp}$. The 
 effective Hamiltonian of N trapped ions is then~\cite{MOR1,MOR2}
 \begin{eqnarray}
 H_{eff}&=&\sum_{i=1}^{N}\{\frac{{\bf p}^2_i}{2}+\frac{1}{2}[ 
           \omega_{\perp}^2(x_i^2+y_i^2)+\omega_{\parallel}^2z_i^2] 
           \\ \nonumber
         &&+U_0cos(\frac{2\pi}{\lambda}z_i)+\sum_{j\ne i}
           \frac{1}{|{\bf r}_i-{\bf r}_j|}\},
 \end{eqnarray}
 where the sinusoidal term represents a laser induced periodic potential,
 mimicking the corrugation of a hypothetical crystalline substrate lattice.\\
 The ground state geometry of the ions at T=0 depends on the aspect ratio
 $R$=$(\omega_{\parallel}/\omega_{\perp})^2$ of the anisotropic harmonic 
 confining potential. For small enough $R$ the potential is cigar-like, and 
 ions form a linear chain along the trap symmetry axis, $z$. As $R$ is 
 increased there is a sequence of shape transitions: first from a straight 
 chain to a planar zigzag chain; next, a second transition where planarity
 is lost, and the planar zigzag turns into a helix. Still at T=0,
 and for an infinite chain, both classical transitions are 
 continuous~\footnote{Note that strictly speaking at T=0 we should not ignore 
 quantum effects and treat this transition as a quantum critical point, as 
 developed for example in Ref.~[\onlinecite{SHIM}]} as shown in Fig.~\ref{fig3}.

\begin{figure}[!t]
 \begin{center}
 \includegraphics[angle=0, width=0.5\textwidth]{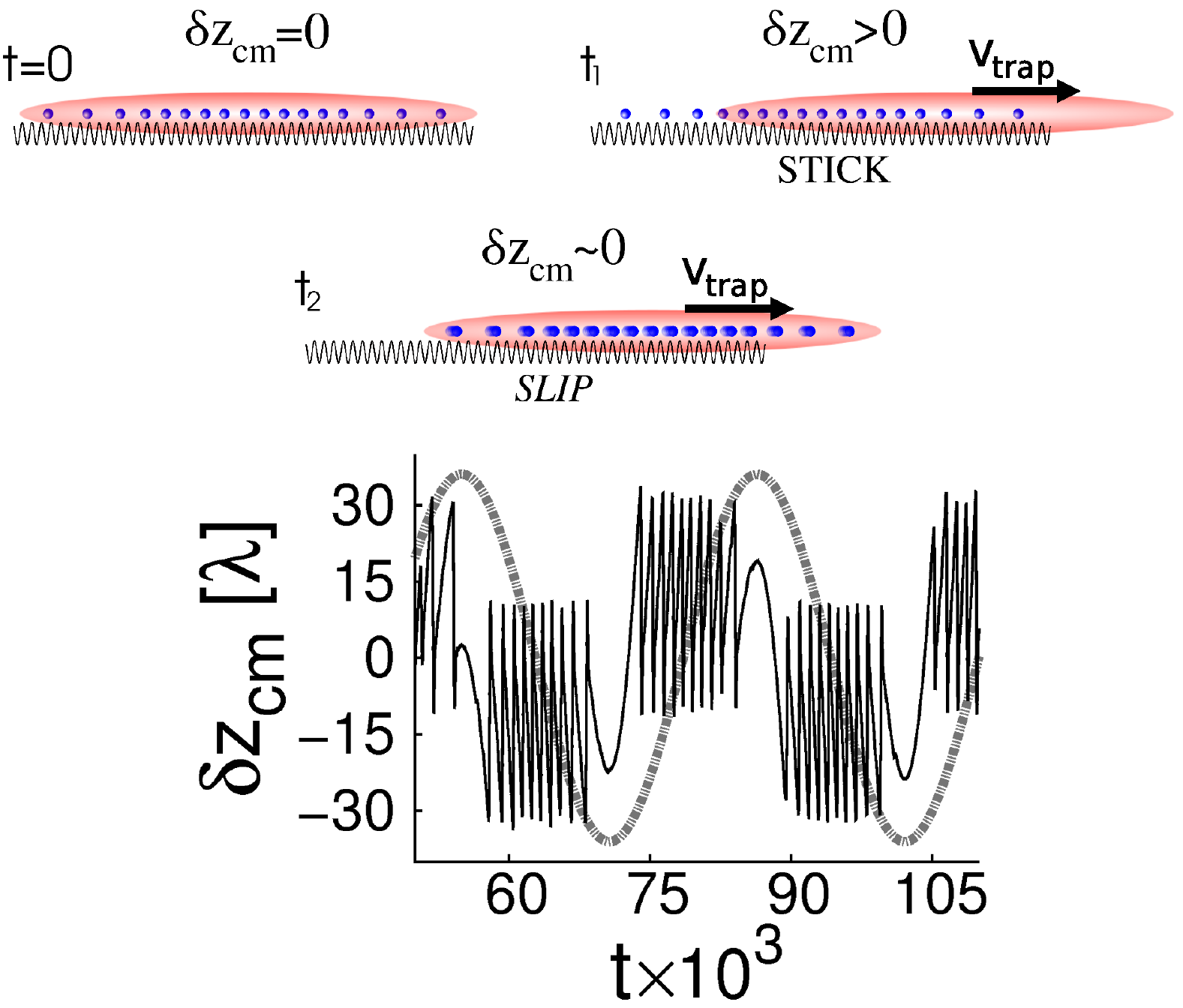}
 \end{center}
 \caption{Dynamics of $\delta z_{cm}$ (see text) under periodic sliding.
          At t=0 the chain is almost symmetric with respect
          to the vertex of the confining parabola and $\delta z_{cm}\approx0$;
          if the chain is locked to the substrate $\delta z_{cm}$ increases
          as the confining potential moves until a slip event occurs, which
          corresponds to a sudden drop to zero of $\delta z_{cm}$. The plot 
          shows a result from a simulation where many slip events are
          observed during each oscillation back and forth of the chain over 
          the substrate. The grey dashed line corresponds to the external 
          electric field.}
 \label{fig4}
 \end{figure}

 We simulated, following Benassi et al.~\cite{BEN}, a chain of 101 positive
 ions choosing $\omega_{\parallel}^2$=0.0005, $R$=0.0005 and 
 $\eta$=$\lambda/a_o$=$2/(1+\sqrt{5})$, where $a_o$ is the center ion-ion 
 spacing. Chain sliding is caused by an external slowly oscillating electric 
 field ${\bf E}(t)$=${\bf \hat{z}}E_0sin(\Omega t)$ acting on each ion in the 
 longitudinal direction $z$. We carried out classical damped molecular dynamics 
 (MD) integrating the equations of motion using a standard velocity-Verlet 
 algorithm with time step $\Delta t$=0.005. At each time the total force 
 acting on the $i^{th}$ ion is given by:
 \begin{equation}
  \ddot{{\bf r}}_i={\bf F}^{Coul}_i+{\bf F}^{trap}_i+{\bf F}^{sub}_i
                   -\gamma {\bf v}_i,
 \end{equation}
 where we have respectively the force due to the ion-ion Coulomb repulsion, 
 the confining potential force, the corrugation force and a velocity dependent
 dissipative force controlled by a damping parameter $\gamma$. There is no 
 random force, corresponding to our ``T=0'' background assumption. The trap 
 confinement plus oscillating potential is given by
 \begin{equation}
 V_{ext}(z)=\frac{\omega_{\parallel}^2}{2}z^2-zE_0sin(\Omega t),
 \end{equation}
 which is a confining parabola of vertex 
 $z_{trap}$=$E_0sin(\Omega t)/\omega_{\parallel}^2$ moving 
 at velocity $v_{trap}$=$E_0\Omega cos(\Omega t)/\omega_{\parallel}^2$. 
 In order to follow stick-slip, when present, we monitored the distance
 of the center of mass of the chain from the minimum of the moving parabola:
 \begin{equation}
 \delta z_{cm}(t)=z_{trap}(t)-z_{cm}(t)
 \end{equation}
 Figure~\ref{fig4} shows an example of the time evolution of $\delta z_{cm}$
 corresponding to a sequence of external electric field oscillations. The
 dynamic friction of the system is computed as the work $W$ done by the
 oscillating electric field on all the particles:
 \begin{equation}
 W_{k}=\frac{1}{N}\sum_{i=1}^{N}\int_{k}dt[\hspace{1pt}{\bf v}_i
       \cdot{\bf \hat{z}}E_0sin(\Omega t)].
 \end{equation}
 Where the integral is calculated over the $k^{th}$ period corresponding
 to the electric field going from its minimum value $-E_0$ to its maximum
 value $+E_0$ and back. The final estimate of the dynamic friction is obtained
 from the average of the M samples $W_k$ measured during the whole trajectory:
 $W$=$\frac{1}{M}\sum_{k=1}^{M}W_k$.
 \begin{figure}[!t]
 \begin{center}
 \includegraphics[angle=0, width=0.5\textwidth]{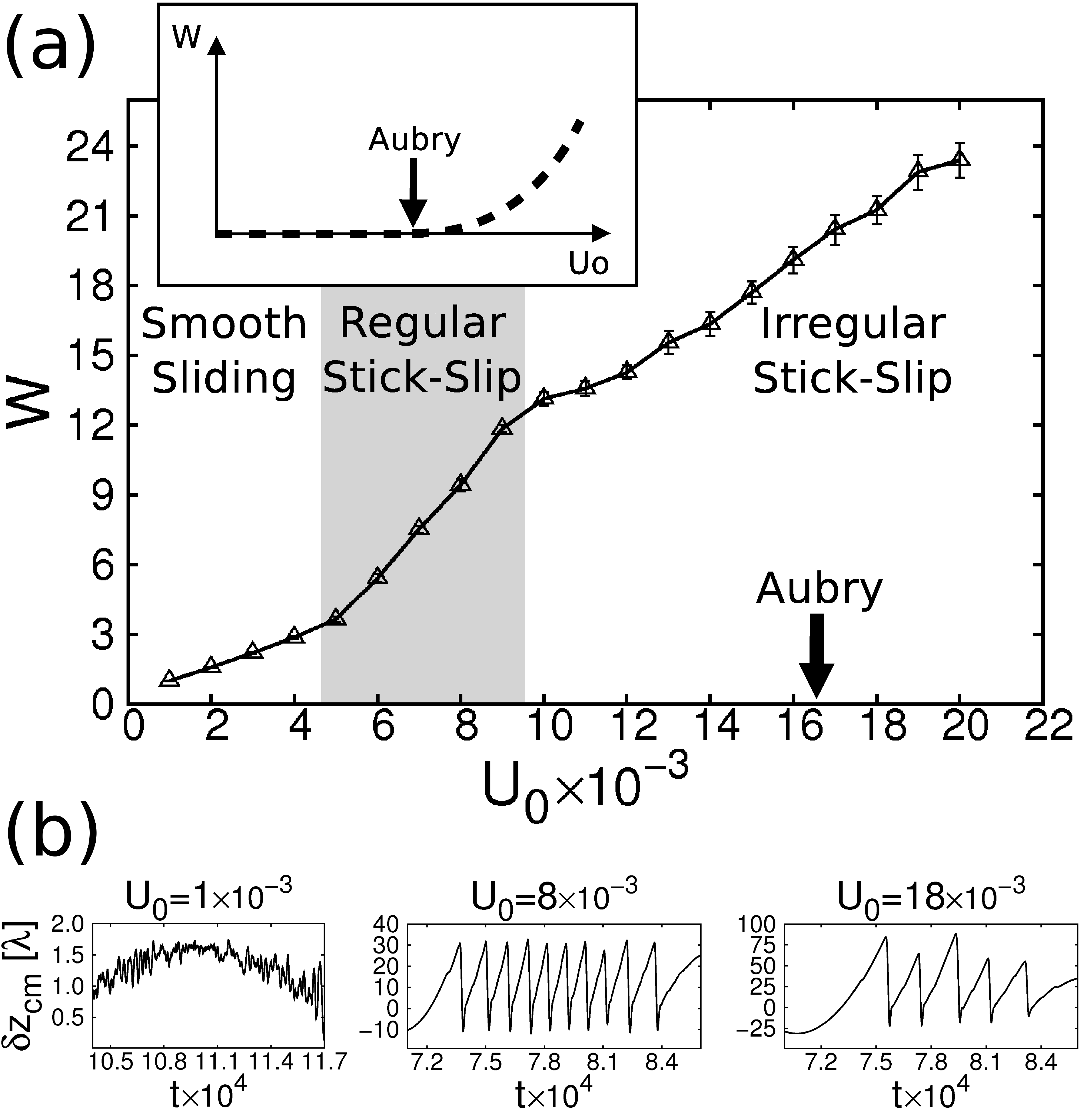}
 \end{center}
 \caption{(a) Dynamic friction $W$ of the oscillating ion chain as a function of
          the corrugation amplitude $U_0$. The values of $E_0$=0.1625,
          $\Omega$=0.0002, $\gamma$=0.01 were chosen so as to yield 4-5
          slip events during each oscillation in correspondence of the highest
          values of $U_0$ investigated. (b) Dynamics of $\delta z_{cm}$ 
          in the three different regimes of smooth-sliding, regular and 
          irregular stick-slip. The time interval plotted corresponds 
          approximately to half a period of the oscillating electric field 
          which increases from $-E_0$ to $+E_0$.}
 \label{fig5}
 \end{figure}
\section{Results: Dynamic Friction, Stick-Slip and Precursor Events}
\label{sec2}
 Figure~\ref{fig5}a shows the frictional work $W$ done by the external electric
 field on the trapped ion chain as a function of the corrugation amplitude 
 $U_0$. The inset depicts schematically the expected behavior in the infinite
 incommensurate FK model, where for $U_0 < U_c$, below the Aubry transition,
 motion takes place without static friction, and kinetic friction vanishes in
 the limit of infinitely slow sliding. In the chain of trapped ions, finite
 and inhomogeneous, the static friction force needed for overall chain motion
 is nonzero for all corrugations, since the two extremities are always locked
 to the corrugation potential. Correspondingly, there is upon sliding a finite
 frictional dissipation $W$ for all values of $U_0$. For the chosen external
 field frequency $\Omega$=0.0002 and amplitude $E_0$=0.1625 friction grows
 steadily with corrugation $U_0$ and nothing significant happens to the dynamic
 friction $W$ across the nominal~\cite{BEN} Aubry value $U_c$=0.01628. In fact
 the dynamics of the chain does not change appreciably above $U_0\approx0.01$.

 We can resolve, based on the detailed nature of the sliding trajectories,
 three different dynamical regimes. Figure~\ref{fig5}b shows
 the dynamics of $\delta z_{cm}$ for three representative values of $U_0$.
 For small corrugations $U_0\leqslant U_{01}$=0.005 the chain
 follows smoothly the external force and the friction is modest (with a value
 determined by, and growing with, the sliding velocity, in turn proportional
 to $\Omega$). As $U_0$ is increased further, the smooth sliding dynamics is
 replaced by a regular, time-periodic stick-slip regime with accompanying 
 increase of dissipation. So long as $U_{01}\leqslant U_0 \leqslant 
 U_{02}$=0.01 the slip magnitude is fairly constant during each oscillation.
 For larger corrugations finally, $U_0\geqslant U_{02}$ the chain enters a
 chaotic regime of irregular stick-slip.

\begin{figure*}[!t]
 \begin{center}
 \includegraphics[angle=0, width=1.0\textwidth]{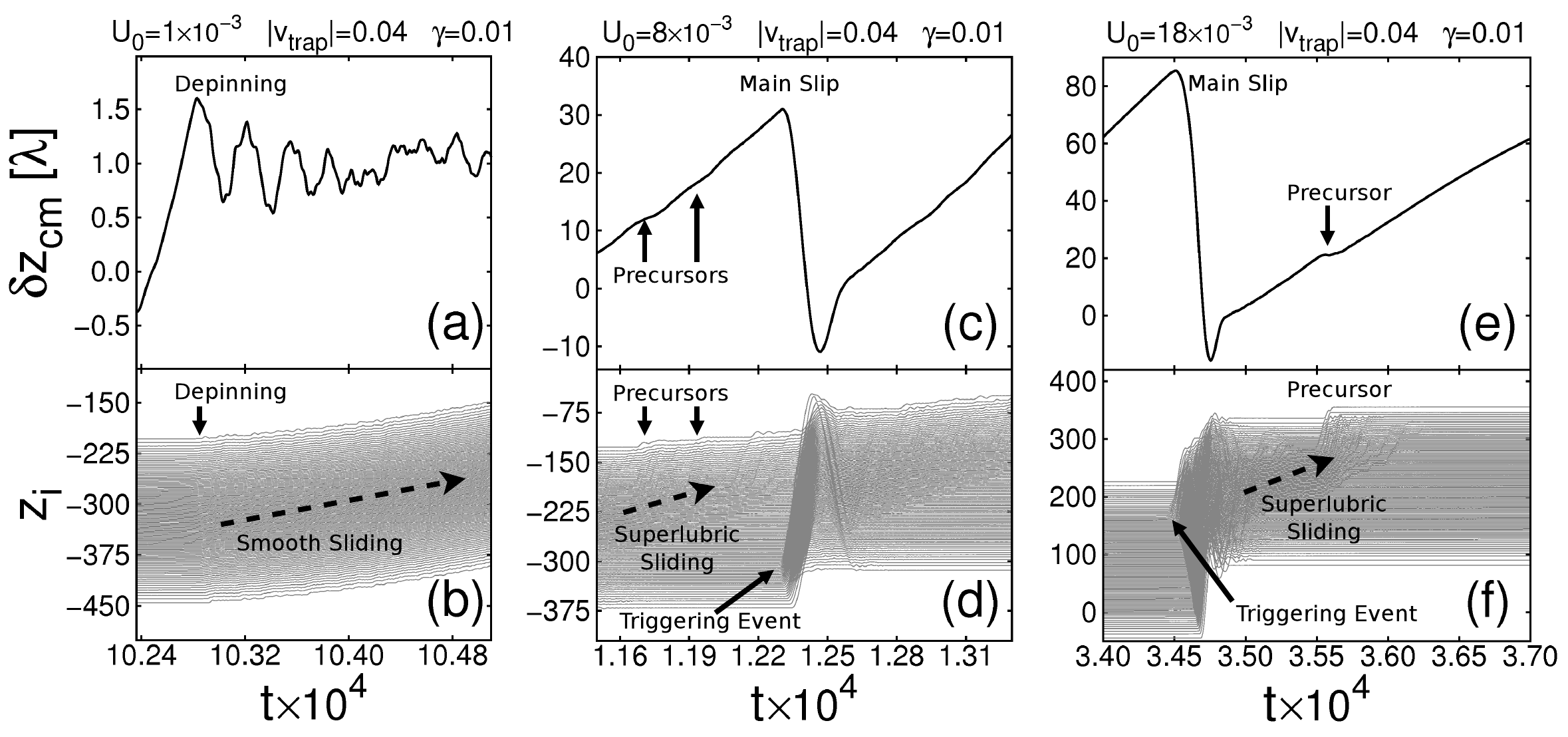}
 \end{center}
 \caption{(a) Dynamics of $\delta z_{cm}$ in the smooth sliding
              regime; panel (b) shows the corresponding trajectories of 
              each ion.
          (c) Dynamics of $\delta z_{cm}$ in the regular stick-slip
              regime; panel (d) shows the corresponding trajectories of each 
              ion.
          (e) Dynamics of $\delta z_{cm}$ in the irregular stick-slip
              regime; panel (f) shows the corresponding trajectories of each 
              ion.\\
          The features highlighted in the figure are discussed in 
          Section~\ref{sec2}.}
 \label{fig6}
 \end{figure*}

\begin{figure}[!t]
 \begin{center}
 \includegraphics[angle=0, width=0.5\textwidth]{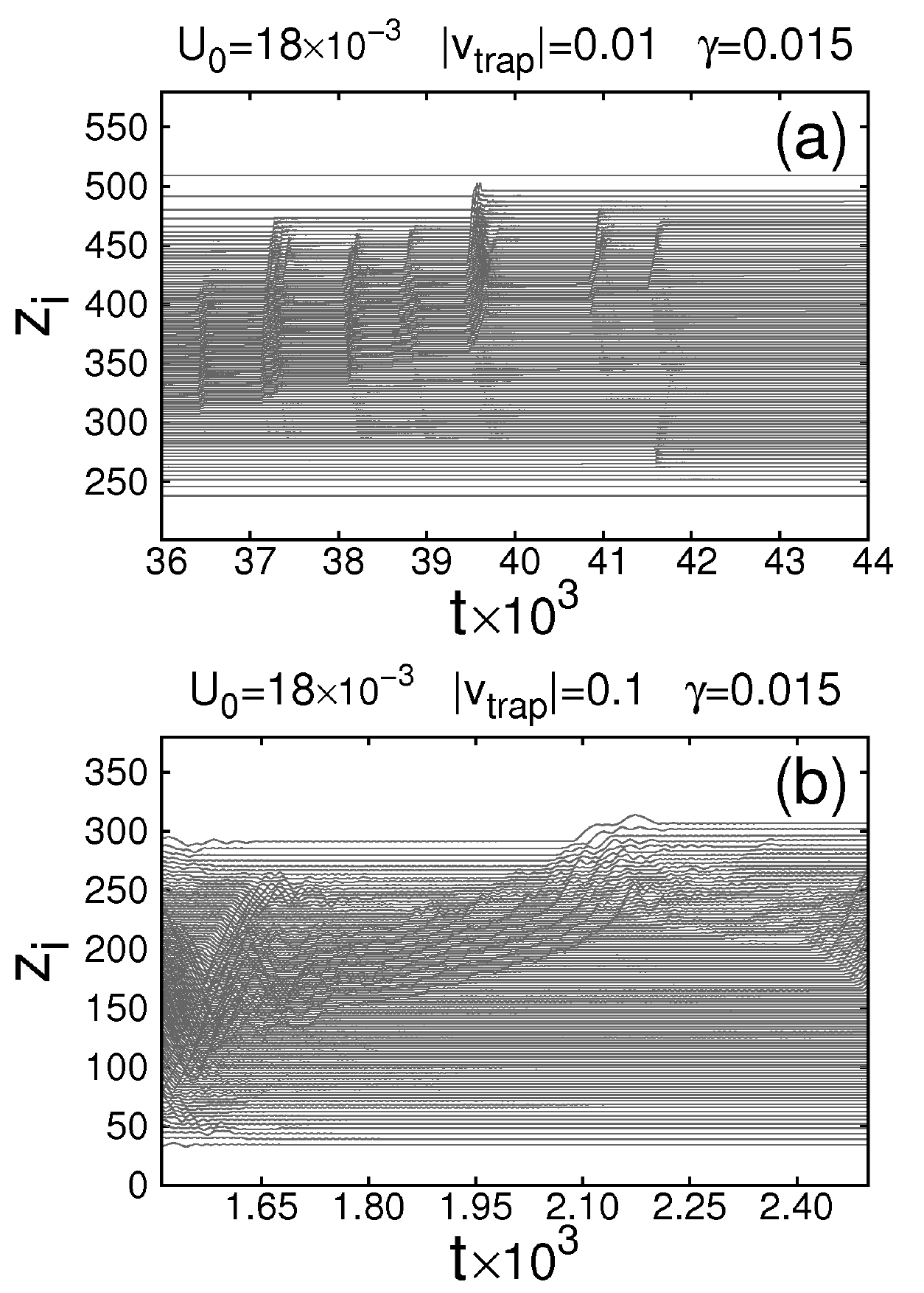}
 \end{center}
 \caption{Precursor events observed between two main slip events for (a)
          $|v_{trap}|$=$10^{-2}$ and (b) $|v_{trap}|$=$10^{-1}$. As the pulling
          velocity is increased the precursor events disappear signaling the
          transition from stick-slip motion to smooth sliding.}
 \label{fig7}

 \end{figure}

\begin{figure}[!t]
 \begin{center}
 \includegraphics[angle=0, width=0.5\textwidth]{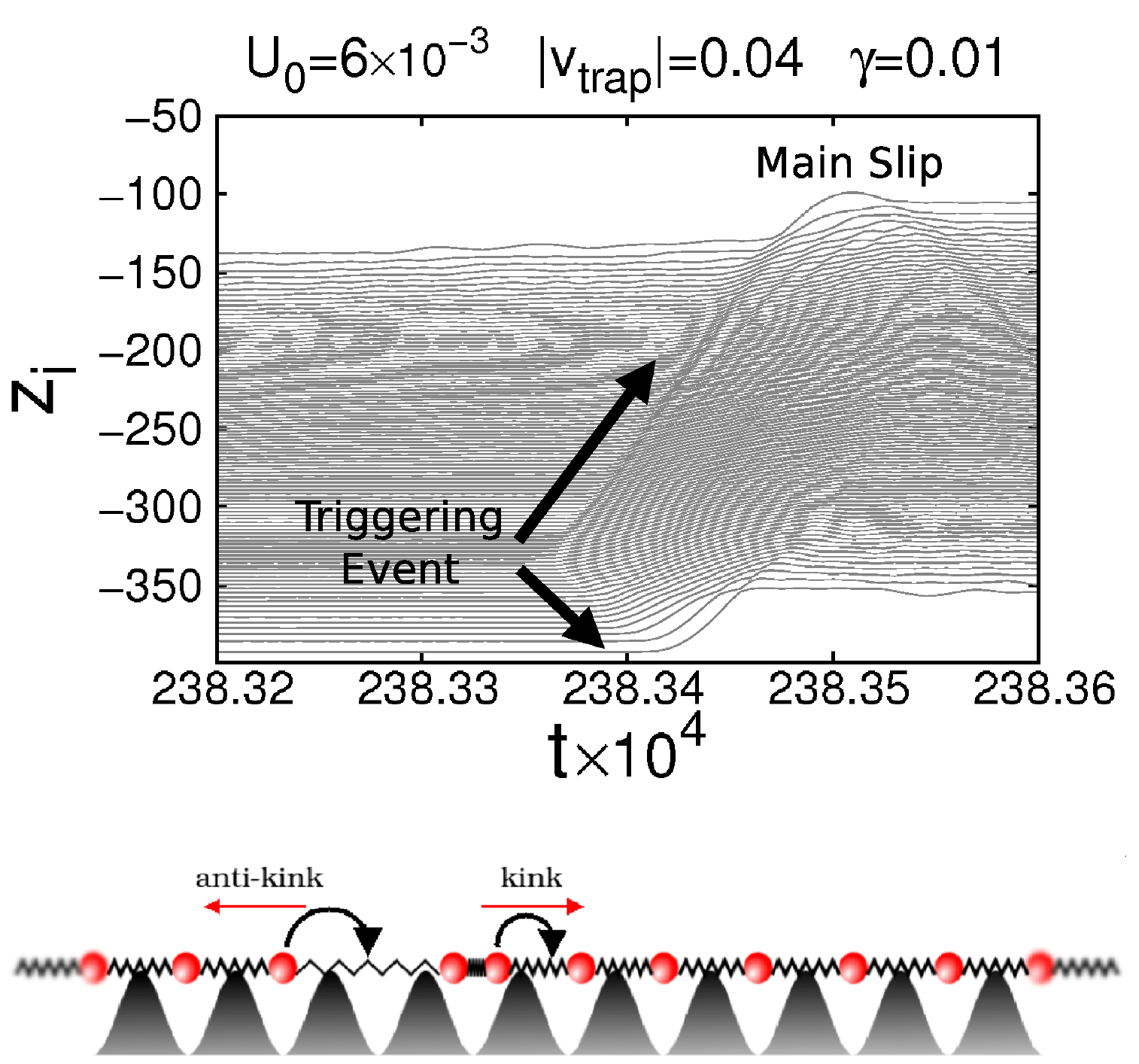}
 \end{center}
 \caption{Example of a triggering event inducing the main slip.}
 \label{fig8}
 \end{figure}

\begin{figure}[!t]
 \begin{center}
 \includegraphics[angle=0, width=0.5\textwidth]{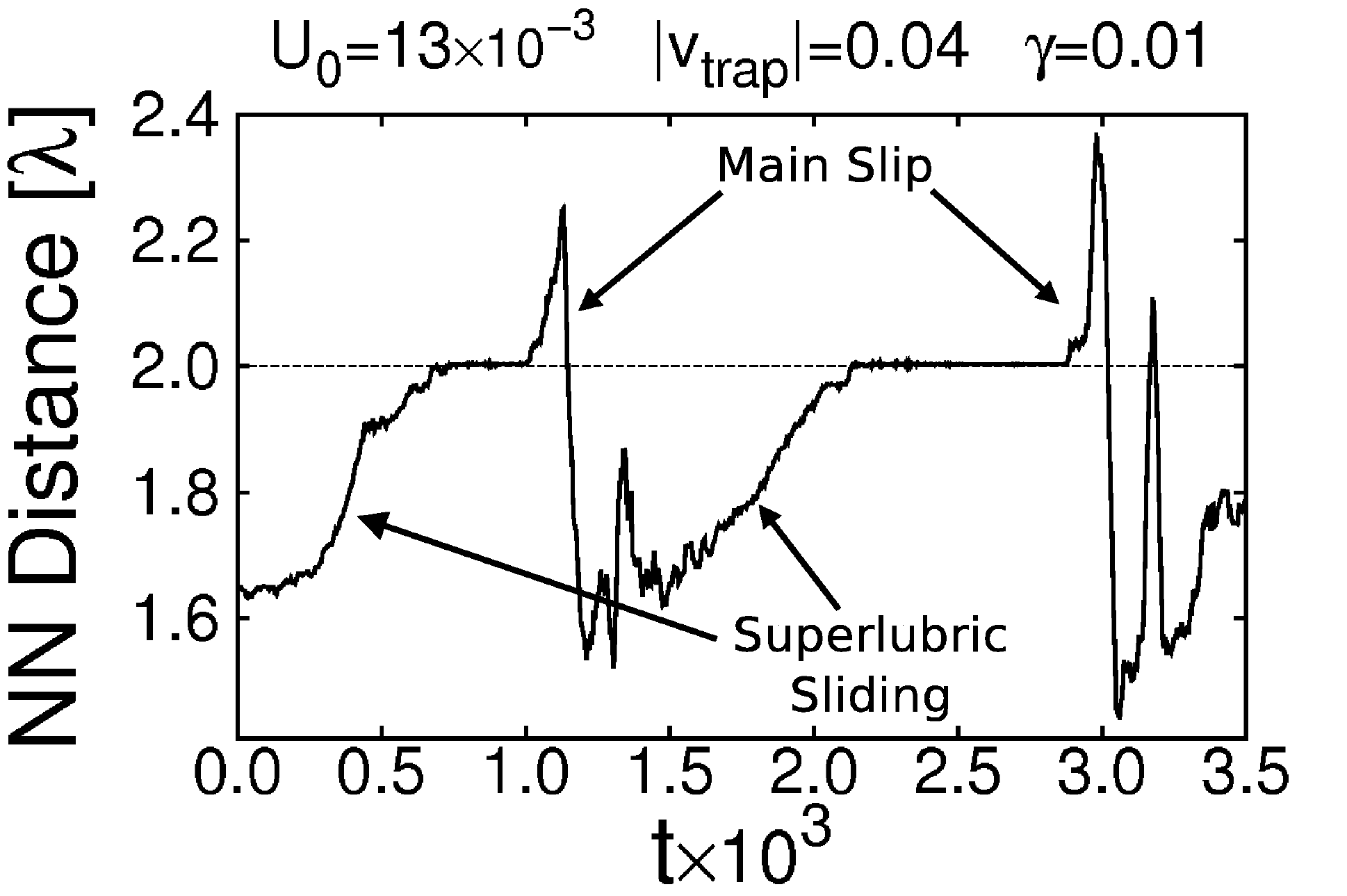}
 \end{center}
 \caption{Average ion-ion distance (in units of $\lambda$) versus time; we
          considered only the central 31 ions of the chain. After the passage
          of the superlubric front the center of the chain is left in a
          commensurate configuration with ions equally spaced by 2 $\lambda$.
          The same kind of dynamics occurs at the onset of motion at each
          values of $U_0$ in the stick-slip region.}
 \label{fig9}
 \end{figure}

 Figure~\ref{fig6} shows details of the trajectories of the chain center of 
 mass and of all individual ions in the three regimes. For $U_0\leqslant 
 U_{01}$ the chain is weakly pinned at the inversions of motion
 occurring for E(t)=$\pm E_0$. After the depinning the chain follows smoothly
 the external force; small oscillations of $\delta z_{cm}$ are due to internal
 motion of the chain (see Fig.~\ref{fig6}a,b).
 A number of interesting features appear at the onset of stick-slip
 $U_0\geqslant U_{01}$ and they are shown in Fig.~\ref{fig6}c-f.
 The head and the tail of the ion crystal are locked to the corrugation
 potential (see top and bottom parts of Fig.~\ref{fig6}d, initial times) 
 while the truly incommensurate central part is free to slide,
 thus increasing the ion density of the head, reducing that of the tail and
 inducing partial depinnings of the chain. The precursor events appear in
 the head part (see top part of Fig.~\ref{fig6}d, times between t=1.16 and
 t=1.22). Although different and connected with inhomogeneity of stress rather
 than of contact, partial precursors were also shown to precede the onset of
 macroscopic sliding by Fineberg's group~\cite{JAY1,JAY2}.
 Partial slips of the chain always start within the central
 superlubric region and proceed moving in the direction of the external force.
 They are present only in the stick-slip regime and disappear by increasing
 the average pulling velocity $|v_{trap}|$=$2E_0\Omega
 /\pi\omega_{\parallel}^2$ of the trapping potential (see Fig.~\ref{fig7}).
 Following the precursors, as stress accumulates more and more 
 in time, the system undergoes a mechanical instability typical of
 stick-slip~\cite{VANRMP}. The ensuing main slip of the whole system is 
 triggered by the creation of a kink-antikink pair (see Fig.~\ref{fig8}). 
 In the regular stick-slip regime $U_{01} \leqslant U_0\leqslant U_{02}$ 
 the tail drives the sliding onset, and the triggering pair forms in the tail
 region, while the chain center and front are still free to slide (see
 Fig.~\ref{fig6}d, times between t=1.22 and t=1.25). In the chaotic stick-slip
 regime $U_0\geqslant U_{02}$ instead, the onset of sliding is different. The
 central superlubric flow and partial slips of the front ions first bring the
 chain into a metastable state where each ion is locked to the corrugation 
 (see Fig.~\ref{fig6}f, times between t=3.5 and t=3.6).
 The generation of a kink-antikink pair forming now in the chain head, as
 opposed to the chain tail of the previous regime, eventually leads to global
 sliding, as shown by the main slip event at the initial times of 
 Fig.~\ref{fig6}f. This tail-to-head switch of the triggering
 event is a characteristic signature always accompanying the passage from
 regular to chaotic stick-slip. On the other hand, neither the sliding onset
 dynamics nor the dynamical friction magnitude finally display any particular
 feature or singularity when the corrugation grows across the Aubry transition.
 This result underlines a substantial difference between the frictional
 behavior of this short and inhomogeneous chain, and that expected of an
 ideally infinite and uniform FK-like chain. In the latter and ideal system
 there is no other transition than Aubry, and in the limit of zero sliding
 speed stick-slip sliding can only take place when static friction turns
 nonzero, which is above the Aubry transition. Figure~\ref{fig9} shows details
 of the sliding dynamics in the strongly corrugated, chaotic stick-slip regime.
 The average nearest neighbor distance between the central 31-ions portion is
 displayed as a function of time between two main slips of the crystal. The
 initial passage of the ``superlubric'' front brings the central part to a
 commensurate configuration with ions spaced exactly by 2 $\lambda$ (instead
 of the original golden ratio spacing 1.618 $\lambda$) from one another. This
 dynamically induced commensuration brings the whole chain to a temporarily
 locked state thus increasing the static friction force needed for the onset
 of overall motion~\footnote{A similar behavior has been observed in MD
 simulations of 2D mesoscopic colloidal monolayers driven over an
 incommensurate optical lattice~\cite{MANI}.}. Subsequent depinning of the
 chain off this locked state only occurs as the external force grows further,
 and is sudden. This two-stage nature of sliding, and the relative abruptness
 of the depinning is at the origin of the chaotic behaviour of stick-slip in
 this regime.
\section{Friction Singularity at the Linear-Zigzag Structural Transition}
\label{sec3}
\begin{figure}[!t]
 \begin{center}
 \includegraphics[angle=0, width=0.5\textwidth]{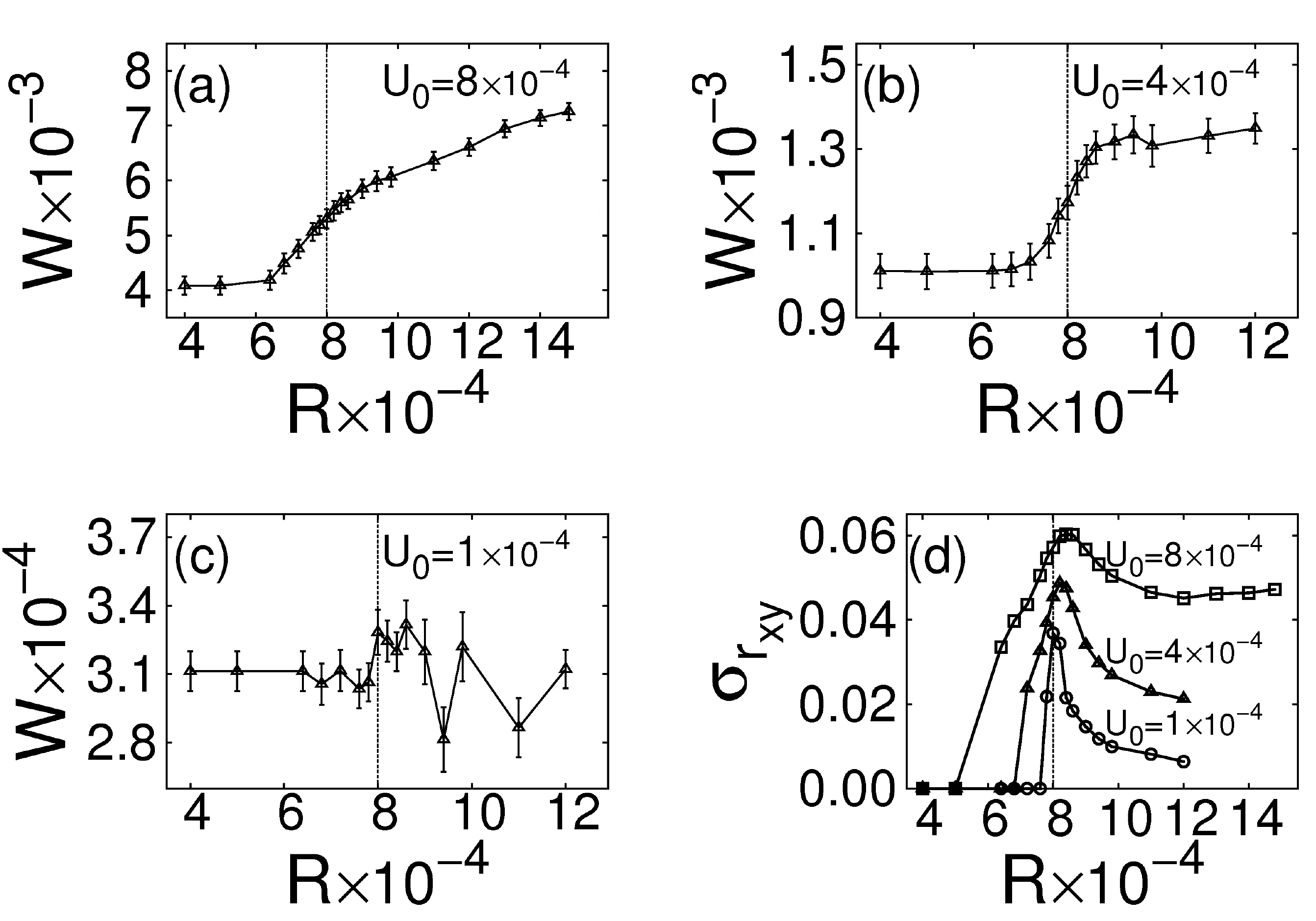}
 \end{center}
 \caption{(a) Dynamic friction across the linear-zigzag transition for
              $U_0$=0.0008. Parameters used are: $\gamma$=0.0005, 
              $E_0$=0.005078125, $\Omega$=0.0032. Without corrugation
              potential the chain moves as a rigid body at a very low 
              dissipation rate independent on $R$ (not shown). As $U_0$ is 
              switched on, the internal degrees of freedom of the chain begin
              to dissipate. In this set of simulations the transverse modes 
              begin to be excited at $R\approx0.00064$. (b) Dynamic friction
              across the linear-zigzag transition for $U_0$=0.0004 and (c)
              $U_0$=0.0001, using $E_0$=0.002539063, $\Omega$=0.0032, 
              $\gamma$=0.0005. By lowering the external electric field and the
              amplitude of the substrate potential the onset of the excitation
              of the transverse modes is shifted towards the static critical
              value $R_c$=0.0008. In this set of simulations the transverse
              modes begin to be excited at $R\approx0.00072$ and 
              $R\approx0.00078$ respectively for $U_0$=0.0004 and $U_0$=0.0001.
          (d) Standard deviation $\sigma_{r_{xy}}$ of the maximum displacement
              $r_{xy}$ of the ions from the $z$ axis plotted as a function of
              $R$ and measured using the whole trajectories of the simulations
              of panels (a),(b),(c). $\sigma_{r_{xy}}$ is zero until the
              transverse modes begin to be excited and it shows a clear maximum
              near $R_c$, which becomes sharper as $U_0$ is decreased.}
 \label{fig10}
 \end{figure}

 Benassi et al. recently proposed that interesting frictional changes or
 anomalies could be observed in presence of structural phase
 transitions~\cite{BEN1}. Although a 101-ions chain is a long way away from
 an infinite system, it is still interesting to find out what singularities
 would friction develop upon an overall, collective shape change. To study
 that, we carried out simulations at values of the trapping potential aspect
 ratio $R$ straddling the critical value $R_c$=0.0008 for the linear-zigzag
 transition. Although this kind of effect should be quite general, the expected
 delicacy of this frictional feature should best become apparent under sliding
 conditions with limited noise, such as those expected at weak corrugations.
 Setting $U_0$=0.0008 and also $\gamma$, $\Omega$ and $E_0$ values which lead
 to a smooth and gentle sliding dynamics (disturbing the chain to a minimal
 extent) we obtained the dynamic friction of Figure~\ref{fig10}a. Deep enough
 in the linear chain regime ($R\leq0.00064$) the ions remains in a strictly
 1D configuration during the whole dynamics. Here, only longitudinal internal
 vibration degrees of freedom of the chain are excited and $W$ is small and
 essentially independent of $R$. As the critical anisotropy $R_c$ is approached
 the transverse vibration modes of the chain soften and become rather suddenly
 excitable, a new dissipative channel opens and $W$ rises anticipating the
 linear-zigzag transition. In a hypothetical infinite chain, where the
 transition occurs continuously and critically, the frictional behavior will
 presumably also exhibit a critical singularity. Given the finite chain size,
 the frictional rise is smooth, although it can still be sharpened by reducing
 the corrugation amplitude. Figures~\ref{fig10}b,c show results of simulations
 with $U_0$ reduced down to 0.0004, 0.0001 and a lower value of $E_0$.
 The transverse mode excitation onset occurs nearer to $R_c$.

 That result is made clearer in Fig.~\ref{fig10}d where the standard deviation
 $\sigma_{r_{xy}}$ of the maximum displacement of the ions away from the $z$
 axis is plotted against the aspect ratio $R$. Denoting with $<\ldots>$ the
 time average over the whole trajectory, $\sigma_{r_{xy}}$ and $r_{xy}$ are
 defined as:
 \begin{equation}
 \sigma_{r_{xy}}=\sqrt{<(r_{xy}-<r_{xy}>)^{2}>}
 \end{equation}
 \begin{equation}
 r_{xy}=\mathrm{Max}_{\{i=1,N_{ions}\}}(\sqrt{x_i^2+y_i^2})
 \end{equation}
 $\sigma_{r_{xy}}$ is zero until the transverse modes begin to be excited and
 it displays a maximum near the critical point $R_c$.

\begin{figure}[!t]
 \begin{center}
 \includegraphics[angle=0, width=0.5\textwidth]{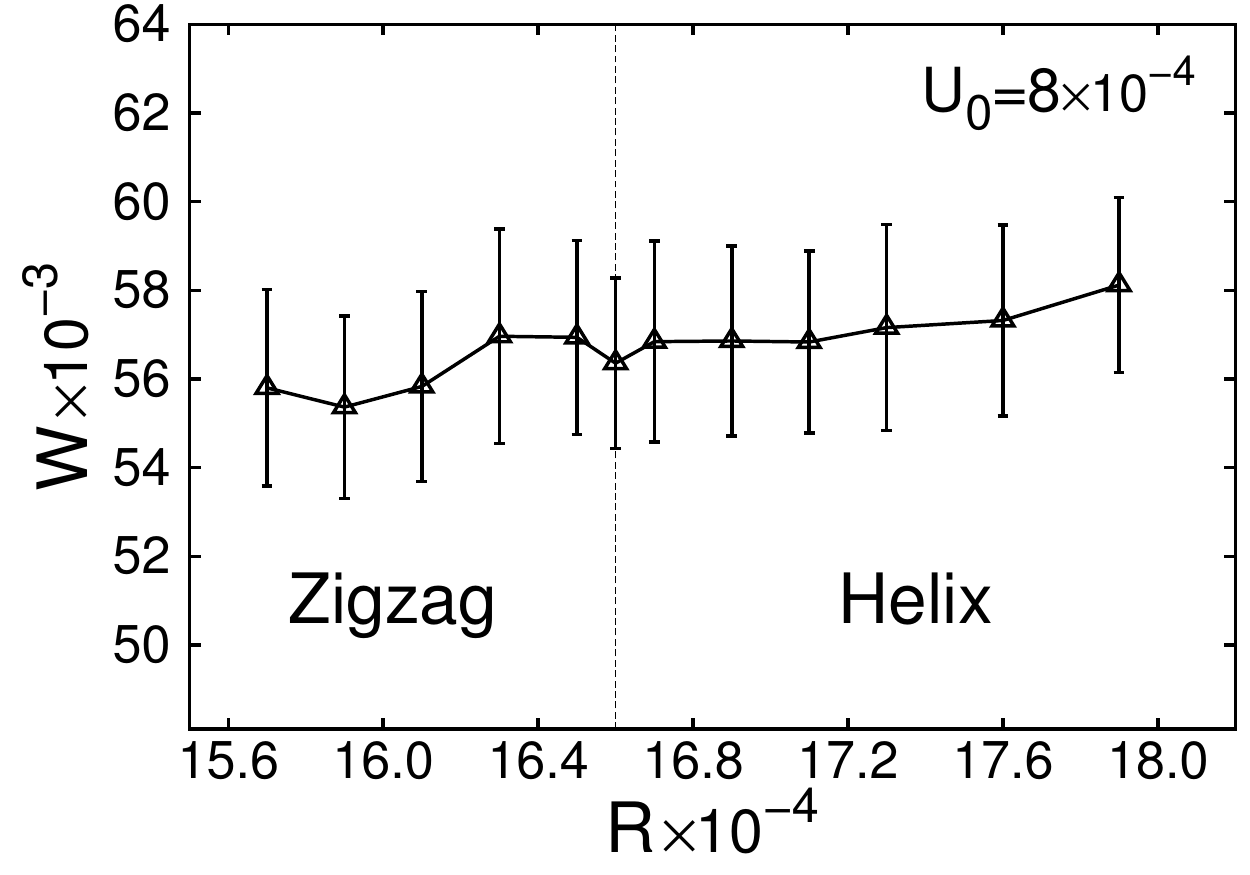}
 \end{center}
 \caption{Dynamic friction across the zigzag-helix transition.
          The dissipative properties of the chain are not affected by the
          helical distortion. Parameters used are: $\gamma$=0.0005,
          $E_0$=0.040625, $\Omega$=0.0004, $U_0$=0.0008.}
 \label{fig11}
 \end{figure}

 A further increase of $R$ drives a subsequent transition of the trapped ion
 geometry from zigzag to helix. Figure~\ref{fig11} shows the behavior which we
 obtained for the dynamic friction across this transition. Not much happens
 here. In this case in fact, on both side of the critical point the chain is
 already in a 3D configuration and its dissipative properties are not
 significantly affected by the weaker helical distortion.
\section{Conclusions}
\label{sec4}
 By means of classical damped MD we simulated a 101-ions linear chain executing
 a forced time-periodic sliding over a (laser induced) space-periodic
 ``corrugation'' potential of strength $U_0$ whose wavelength $\lambda$
 was golden mean incommensurate with respect to the center ion-ion spacing.
 As $U_0$ was increased the system turned from smooth sliding to a stick-slip
 sliding regime, first regular and then chaotic. We observed as expected an
 increase in the dissipation rate at the onset of stick-slip, as is also
 observed in macroscopic dry friction when the loading force is increased.

 Three separate frictional regimes were identified, as a function of
 corrugation amplitude. A smooth sliding one for weak corrugation was followed
 by time-periodic stick-slip sliding at larger corrugation, eventually leading
 to chaotic stick-slip for even larger corrugation.

 Due to the inhomogeneity of the ion crystal the frictional dynamics of the ion
 chain showed several novel features reflected in the ion trajectories. The two
 chain extremities were always pinned while the incommensurate central part was
 free to slide following the external force. The onset of motion in the
 stick-slip regime was characterized by the presence of precursor events, i.e.
 partial slips of side chain portions induced by the superlubric flow of the
 truly incommensurate central part.
 
 The chaotic stick-slip at large corrugation was connected by an interesting
 two-stage process. First, superlubric sliding of the central portion brought
 the chain to a temporary commensurate state, locked to the periodic
 corrugation. Subsequently, as force grew, a kink-antikink pair was generated
 and propagated toward the extremities eventually inducing the slip of the
 whole system.

 We also studied the possible anomalies of friction dissipation across the
 structural phase transformations of the trapped ions obtained by varying the
 aspect ratio $R$=$(\omega_{\parallel}/\omega_{\perp})^2$ 
 of the harmonic trapping potential. At fixed $\omega_{\parallel}$, as $R$ was
 increased, the ion chain transformed first from a linear configuration to a
 planar zigzag and then to a helix. The energy dissipation increased
 characteristically at the linear to zigzag transition. Conversely, the zigzag
 to helix transition did not yield significant frictional changes.

 Considerations about experimentally accessible system parameters, which are
 given in the appendix, suggest that some of these features, if not all,
 should become observable in future experiments with Ca$^+$ ion traps.

\section {Acknowledgments}
\label{sec5}
 This work was partly sponsored by contracts PRIN/COFIN 2010LLKJBX\_004,  
 Sinergia CRSII2$_1$36287/1, and by advances of ERC Advanced Grant 320796 - 
 MODPHYSFRICT. Discussions with A. Benassi, T. Pruttivarasin and H. 
 H\"affner are gratefully acknowledged.
\appendix*
\section{Practical Parameter Choices}
\label{sec6}
 In order to create an optical lattice the laser wavelength must fit one of
 the electronic transition of the chosen ions. For Ca$^+$ the
 S$_{1/2}$-P$_{1/2}$ transition at 397 nm is naturally exploited, leading to a
 lattice spacing $\lambda\approx200$ nm. An achievable amplitude for the
 corrugation potential is of the order of 10$^{-27}$ J. If we consider a
 transverse trapping frequency $\omega_{\perp}$=2$\pi\times$4 MHz we get
 (in dimensionless units) $\lambda$=0.115, U$_0$=2.31$\times$10$^{-5}$.
 A practically achieved temperature is T=10$^{-7}$ corresponding to
 1 $\mathrm{\mu}$K. Setting an aspect ratio $R$=0.0005 we have $a_o\approx$16
 $\lambda$, therefore ions are separated by several lattice spacings,
 corresponding to a much ``weaker'' kind of incommensurability that the
 golden ratio used in the study so far. Moreover for such a small value of
 $U_0$ the chain is almost free to slide, in this case also the two extremities
 being weakly anchored to the substrate. Stick-slip is therefore expected to
 occur only at very small pulling velocities when, after each slip, the chain
 is allowed to relax in a new metastable pinned configuration. We performed
 simulations at the experimental parameters for the 101-ions and a 35-ions
 chain using the same protocol described above. We chose $\gamma$=0.0005 and we
 used a Langevin thermostat for the simulations at finite T. We also set
 $\Omega$ and $E_0$ in order to test different average velocities of the
 moving confining parabola.
 \begin{figure}[!t]
 \begin{center}
 \includegraphics[angle=0, width=0.5\textwidth]{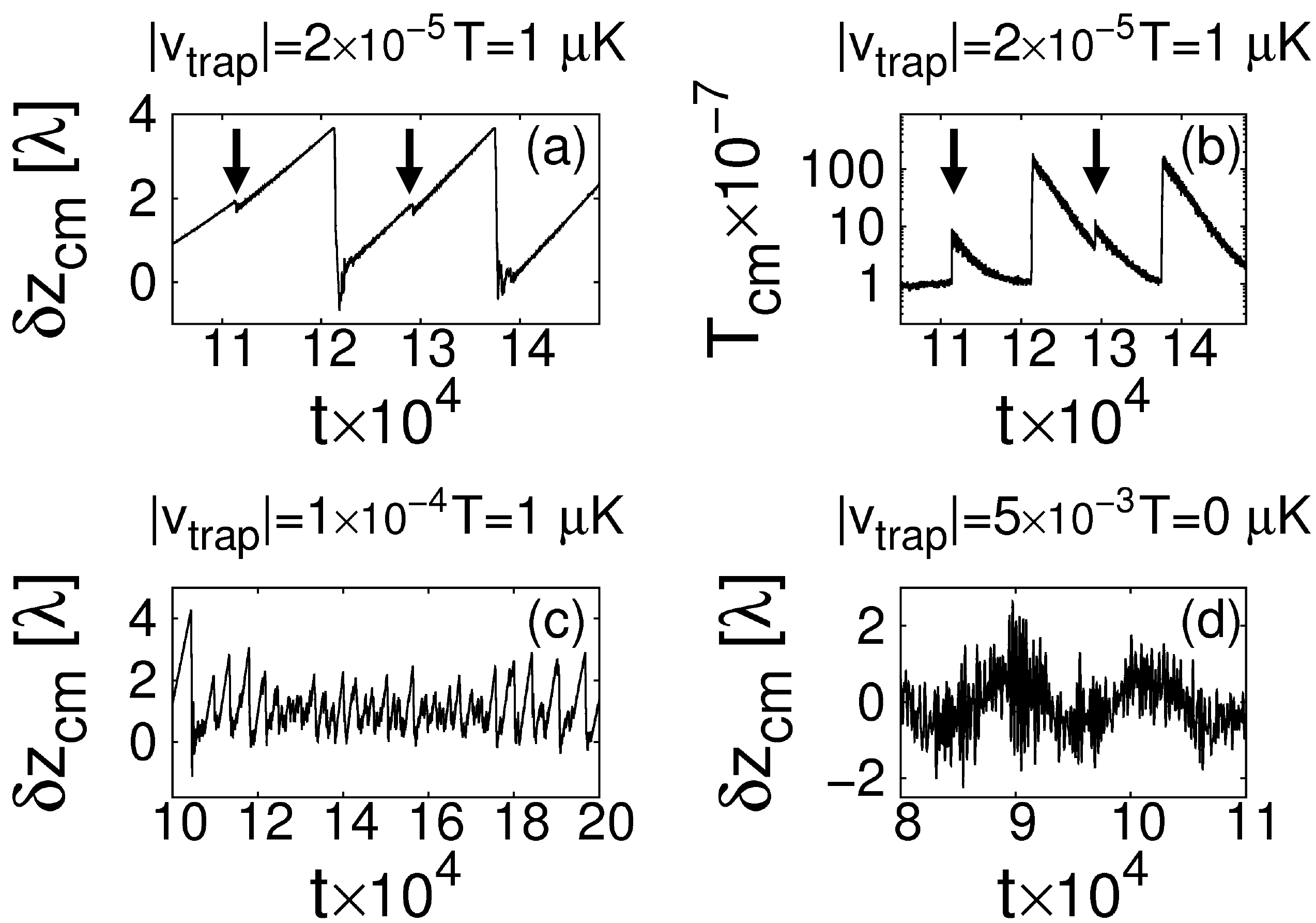}
 \end{center}
 \caption{101-ions chain.\\
         (a) $\delta z_{cm}$ in the simulation at
             $|v_{trap}|$=2$\times$10$^{-5}$ and T=10$^{-7}$ showing stick-slip
             and precursor events (indicated by the arrows) corresponding to
             partial slips of the central portion of the chain.
         (b) Temperature measured in the center of mass frame in the simulation
             of panel (a). After each slip event the internal temperature is
             raised and then it is exponentially damped by the thermostat.
             Precursor events give rise to the smaller peaks indicated by the
             arrows. This trend is always observed in the stick-slip regime.
         (c) For $|v_{trap}|$=10$^{-4}$ and T=10$^{-7}$ the dynamics is chaotic.
             Slips of different magnitude occur as the pulling velocity changes
             during each oscillation. 
         (d) Further increasing $|v_{trap}|$=5$\times10^{-3}$ the dynamics 
             finally turns into a smooth-sliding regime, even at T=0.}
 \label{fig12}
 \end{figure}

\begin{figure}[!t]
 \begin{center}
 \includegraphics[angle=0, width=0.5\textwidth]{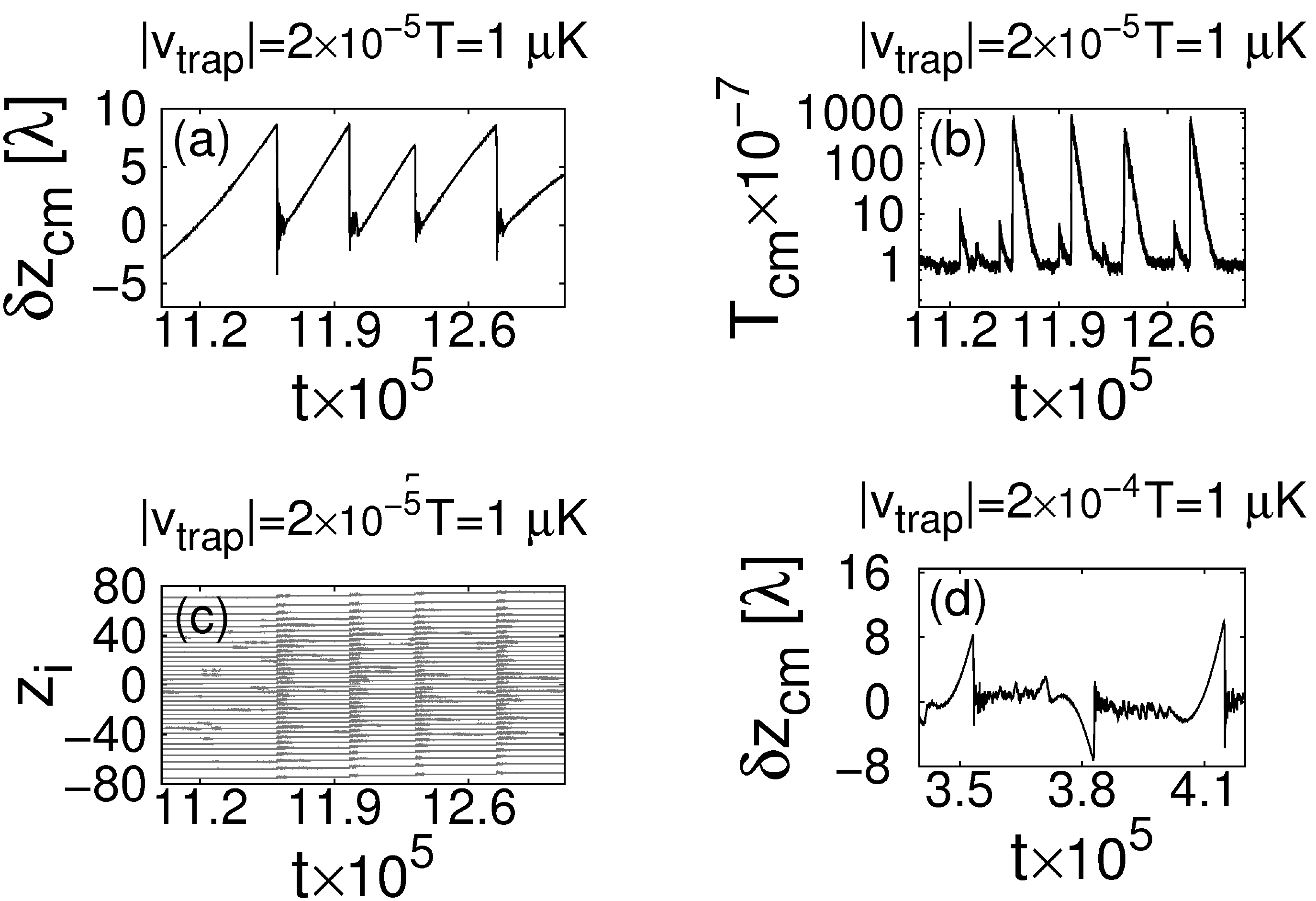}
 \end{center}
 \caption{35-ions chain.\\
          (a)$\delta z_{cm}$ in the simulation at
             $|v_{trap}|$=2$\times$10$^{-5}$ and T=10$^{-7}$. An
             irregular stick-slip regime is observed.
          (b)Temperature measured in the center of mass frame in the
             simulation of panel (a). The highest peaks correspond to the main
             slip events, slip of few ions within the chain give rise to the
             smaller ones.
          (c)Trajectory of each ions during some slip events of the simulation
             of panels (a) and (b). The chain slip as a whole and
             precursor events are not observed due to the small size of the
             system.          
          (d)$\delta z_{cm}$ in the simulation at
             $|v_{trap}|$=2$\times$10$^{-4}$ and T=10$^{-7}$. At the 
             inversions of motion the chain is pinned; owing to the small
             initial pulling velocity a sharp slip event occurs after which
             smooth sliding begins.}
 \label{fig13}
 \end{figure}

 Let us consider first the 101-ions chain.
 At very low pulling velocity we observed stick-slip motion, the slip 
 amplitude being of the order of a few lattice parameters (see Fig.~\ref{fig12}a).
 The dynamics of the slip events is simple and no longer shows the variety of
 features described previously for the golden ratio incommensurability.
 The pinned chain remained stable upon pulling until a weak compression
 generated within the chain propagates toward the head. Simple precursor
 events are observed consisting of partial slips of a central portion of the
 chain at fixed extremities (not shown). Figure~\ref{fig12}b shows the 
 temperature T$_{cm}$ measured in the center of mass frame, displaying the
 expected inverse sawtooth behavior when plotted in a semilogarithmic scale.
 After each slip T$_{cm}$ increases, and is then exponentially damped by the
 thermostat. As the pulling velocity is increased the dynamics turns gradually
 into a smooth-sliding regime, as shown in Fig.~\ref{fig12}c,d.\\
 Chains of a few tens of ions may be more easily stabilized inside a trap. We
 performed simulations with a 35-ions chain, using the same parameters as above.
 In this case the central ion-ion spacing is larger and corresponds to
 $a_o\approx$30 $\lambda$. As shown in Fig.~\ref{fig13}a we observed that
 stick-slip motion is again preserved in this case as well for small enough
 average pulling velocities. The slip amplitude is larger than that observed
 for the 101-ions chain indicating a stronger pinning to the substrate, as is
 reasonable to expect given a larger prevalence of extremities. From the plot
 of the temperature of Fig.~\ref{fig13}b we see that small adjustments of the
 inner ions occur, which give rise to small peaks in T$_{cm}$, however without
 proper precursor events due to the reduced size of the chain.

%

\nocite{SHIM}
\bibliography{BC12526}

\end{document}